\documentclass[%
 aip,
rsi,%
 amsmath,amssymb,
 reprint,%
]{revtex4-1}

\usepackage{graphicx}
\usepackage{dcolumn}
\usepackage{bm}

\usepackage{xcolor}

\usepackage{todonotes}

\begin{document}

\preprint{AIP/123-QED}

\title[MSM estimation with short reseeding trajectories]{Adaptive Markov State Model estimation using short reseeding trajectories}




\author{Hongbin Wan}
\author{Vincent A. Voelz}
\email{voelz@temple.edu}
\affiliation{Department of Chemistry, Temple University, Philadelphia, PA 19122, USA}

\date{\today}

\begin{abstract}
In the last decade, advances in molecular dynamics (MD) and Markov State Model (MSM) methodologies have made possible accurate and efficient estimation of kinetic rates and reactive pathways for complex biomolecular dynamics occurring on slow timescales.  A promising approach to enhanced sampling of MSMs is to use so-called “adaptive” methods, in which new MD trajectories are “seeded” preferentially from previously identified states.  Here, we investigate the performance of various MSM estimators applied to reseeding trajectory data, for both a simple 1D free energy landscape, and for mini-protein folding MSMs of WW domain and NTL9(1-39).  Our results reveal the practical challenges of reseeding simulations, and suggest a simple way to reweight seeding trajectory data to better estimate both thermodynamic and kinetic quantities.
\end{abstract}

\maketitle


\section*{Introduction}
In the last decade, Markov State Model (MSM) methodologies have made possible accurate and efficient estimation of kinetic rates and reactive pathways for slow and complex biomolecular dynamics.\cite{Noe:2009en,Voelz:2010hs,Prinz:2011id,Chodera:2014gk,noe2014introduction} One of the key advantages touted by MSM methods is the ability to use large ensembles of short-timescale trajectories for sampling events that occur on slow timescales.  The main idea is that sufficient sampling using many short trajectories can circumvent the need to sample long trajectories.

With this in mind, many ``adaptive'' methods have been developed for the purpose of accelerating sampling of MSMs.  The simplest of these can be called \textit{adaptive seeding}, where one or more new rounds of unbiased simulations are performed by ``seeding'' swarms of trajectories throughout the landscape.\cite{Huang:2009bx} The choice of seeds are based on some initial approximation of the free energy landscape, possibly from non-equilibrium or enhanced-sampling methods.  Adaptive seeding can be performed by first identifying a set of metastable states, then initiating simulations from each state.  If the seeding trajectories provide sufficient connectivity and statistical sampling of transition rates, an MSM can be constructed to accurately estimate both kinetics and thermodynamics.  
 
Similarly, so-called \textit{adaptive sampling} algorithms have been developed for MSMs in which successive rounds of targeted seeding simulations are performed, updating the MSM after each round.\cite{Voelz:2014kk,Shamsi:2017jg} A simple adaptive sampling strategy is to start successive rounds of simulations from under-sampled states, for instance, from the state with the least number of transition counts.\cite{Doerr:2014fc}  A more sophisticated approach is the FAST algorithm, which is designed to discover states and reactive pathways of interest by choosing new states based on an objective function that balances under-sampling with a reward for sampling desired structural observables.\cite{Zimmerman:2015kf,Zimmerman:2018jn} Other algorithms include: REAP, which efficiently explores folding landscapes by using reinforcement learning to choose new states,\cite{shamsi2017reinforcement} and surprisal-based sampling,\cite{Voelz:2014kk} which chooses new states that will minimize the uncertainty of the relative entropy between two or more MSMs.

A key problem with adaptive sampling of MSMs arises because we are often interested in \textit{equilibrium} properties, while trajectory seeding is decidedly \textit{non-equilibrium}.  This may seem like a subtle point, because the dynamical trajectories themselves are unbiased, but of course, the ensemble of starting points for each trajectory are almost always \textit{statistically} biased, i.e. the seeds are not drawn from the true equilibrium distribution.  This can be problematic because most MSMs are constructed from transition rate estimators that enforce detailed balance and assume trajectory data is obtained at equilibrium. The distribution of sampled transitions, however, will only reflect equilibrium conditions in the limit of long trajectory length.

One way around this problem is to focus mostly on the kinetic information obtained by adaptive sampling.  A recent study of the ability of FAST to accurately describe reactive pathways concluded that the most reliable MSM estimator to use with adaptive sampling data is a row-normalized transition count matrix.\cite{Zimmerman:2018jn} Indeed, weighted-ensemble path sampling algorithms focus mainly on sampling the kinetics of reactive pathways, information which can be used to recover global thermodynamic properties.\cite{BinWZhang:2010kf,Zwier:2015fn, Dickson:2016it, Lotz:2018hx, Dixon:2018fs} A major disadvantage of this approach is that it ignores potentially valuable equilibrium information.  As shown by Trendelkamp-Schroer and No{\'e},\cite{trendelkamp2016efficient} detailed balance is a powerful constraint to infer rare-event transition rates from equilibrium populations.  Specifically, when faced with limited sampling, dedicating half of one's simulation samples toward enhanced thermodynamic sampling (e.g. umbrella sampling) can result in a significant reduction in the uncertainty of estimated rates, simply because the improved estimates of equilibrium state populations inform the rate estimates through detailed balance.

Another way around this problem, recently described by N{\"u}ske et al., is to use an estimator based on observable operator model (OOM) theory, which utilizes information from transitions observed at lag times $\tau$ and $2\tau$ to obtain estimates unbiased by the initial distribution of seeding trajectories.\cite{Nuske:2017ex}  Although the OOM estimator is able to make better MSM estimates at shorter lag times, it requires the storage of transition count arrays that scale as the cube of the number states, and dense-matrix singular-value decomposition, which can be impractical for MSMs with large numbers of states.  N{\"u}ske et al. derive an expression quantifying the error incurred by non-equilibrium seeding, from which they conclude that such bias is difficult to remove without either increasing the lag time or improving the state discretization.  

Here, we explore an alternative way to recover accurate MSM estimates from biased seeding trajectories, by reweighting sampled transition counts to better approximate counts that would be observed at equilibrium.  Like the Trendelkamp-Schroer and No{\'e} method,\cite{trendelkamp2016efficient} this requires some initial estimate of state populations, perhaps obtained from previous rounds of adaptive sampling.

We are particularly interested in examining how the performance of this reweighting method compares to other estimators, in cases where it is impractical to generate long trajectories and instead one must rely on ensembles of short seeding trajectories. An example of a case like this is adaptive seeding of protein folding MSMs built from ultra-long trajectories simulated on the Anton supercomputer.\cite{LindorffLarsen:2011gl} Because such computers are not widely available, adaptive seeding using conventional computers may be one of the only practical ways to leverage MSMs to predict the effect of mutations, for example. 

In this manuscript, we first perform adaptive seeding tests using 1D-potential energy models, and compare how different estimators perform at accurately capturing kinetics and thermodynamics.  We then perform similar tests for MSMs built from  ultra-long reversible folding trajectories of two mini-proteins, WW domain and NTL9(1-39). Our results, described below, suggest that reweighting trajectory counts with estimates of equilibrium state populations can achieve a good balance of kinetic and thermodynamic estimation.

\subsection*{Estimators}

We explored the accuracy and efficiency of several different transition probability estimators using  adaptive seeding trajectory data as input: (1) a maximum-likelihood estimator (MLE), (2) a MLE estimator where the exact equilibrium populations $\pi_i$ of each state $i$ are known \textit{a priori}, (3) an MLE estimator where each input trajectory is weighted by an \textit{a priori} estimate of the equilibrium population of its starting state, (4) row-normalized transition counts, and (5) an observable operator model (OOM) estimator.

\subsubsection*{Maximum-likelihood estimator (MLE).} 
The MLE for a reversible MSM assumes that observed transition counts are independent, and drawn from the equilibrium distribution, so that reversibility (i.e. detailed balance) can be used as a constraint.  The likelihood of observing a set of given transition counts, $L = \prod_i \prod_j p_{ij}^{c_{ij}}$, when maximized under the constraint that $\pi_i p_{ij} = \pi_j p_{ji}$, for all $i, j$, yields a self-consistent expression that can be iterated to find the equilibrium populations,\cite{Wu:2014jya, Prinz:2011id, Bowman:2009jw}
\begin{equation}
\pi_i  = \sum_j \frac{c_{ij} + c_{ji}}{\frac{N_j}{\pi_j} + \frac{N_i}{\pi_i}}
\end{equation}
where $N_i = \sum_j c_{ij}$.  The transition probabilities $p_{ij}$ are  given by 
\begin{equation}
p_{ij} = \frac{(c_{ij} + c_{ji})\pi_j }{N_j\pi_i + N_i\pi_j } 
\end{equation}

\subsubsection*{Maximum-likelihood estimator (MLE) with known populations $\pi_i$}
Maximization \color{black} of the likelihood function above, with the additional constraint of fixed populations $\pi_i$, yields a similar self-consistent equation that can be used to determine a set of Lagrange multipliers,\cite{TrendelkampSchroer:2016bz}
\begin{equation}
\lambda_i  = \sum_j \frac{(c_{ij} + c_{ji})\pi_j \lambda_i}{\lambda_j\pi_i + \lambda_i\pi_j },
\end{equation}
from which the transition probabilities $p_{ij}$ can be obtained as
\begin{equation}
p_{ij} = \frac{(c_{ij} + c_{ji})\pi_j }{\lambda_j\pi_i + \lambda_i\pi_j }.
\end{equation}

\subsubsection*{Maximum-likelihood estimator (MLE) with population-weighted trajectory counts.}
For this estimator, first a modified count matrix $c_{ij}'$ is calculated, 
\begin{equation}
c_{ij}' = \sum_k w^{(k)}c_{ij}^{(k)},
\end{equation}

where transition counts $c_{ji}^{(k)}$ from trajectory $k$ are weighted in proportion to $w^{(k)} = \pi^{(k)}$, the estimated equilibrium population of the initial state of the trajectory.  The idea behind this approach is to counteract the statistical bias from adaptive seeding by scaling the observed transition counts proportional to their equilibrium fluxes. The modified counts are then used as input to the MLE.

\subsubsection*{Row-normalized counts.}
For this estimator, the transition probabilities are approximated as
\begin{equation}
p_{ij} = \frac{c_{ij}}{\sum_j c_{ij}}.
\end{equation}
This approach does not guarantee reversible transition probabilities, which only occurs in the limit of large numbers of reversible transition counts.  In practice, however, the largest eigenvectors of the transition probability matrix have very nearly real eigenvalues, such that we can report relevant relaxation timescales and equilibrium populations. 

\subsubsection*{Observable operator model (OOM) theory}

For an introduction to OOMs and their use in estimating MSMs, see references.\cite{Jaeger:2000ui,Wu:2015jda,Nuske:2017ex} OOMs are closely related to Hidden Markov Models (HMMs) but have the advantage, like MSMs, that they can be learned directly from observable-projected trajectory data.\cite{Wu:2015jda}  Unlike MSMs, they require the collection of both one-step transition count matrices, and a complete set of two-step transition count matrices. With enough training data, and sufficient model rank, OOMs can recapitulate exact relaxation timescales, uncontaminated by MSM state discretization error.\cite{Nuske:2017ex}

We used the OOM estimator as described by N{\"u}ske et al.\cite{Nuske:2017ex} and implemented in PyEMMA 2.3,\cite{Scherer:2015jb} using the default method of choosing the OOM model rank by discarding singular values that contribute a bootstrap-estimated signal-to-noise ratio of less than 10.  The settings used to instantiate the model in PyEMMA are: \texttt{
OOMReweightedMSM(reversible=True, \\ count{\_}mode='sliding', sparse=False, \\ connectivity='largest', rank{\_}Ct='bootstrap{\_}counts', tol{\_}rank=10.0, score{\_}method='VAMP2', score{\_}k=10, mincount{\_}connectivity='1/n')
}

The OOM estimator returns two estimates: (1) a so-called \textit{corrected} MSM, which derives from using the unbiased OOM equilibrium correlation matrices to construct a matrix of MSM transition probabilities, and (2) the \textit{OOM timescales}, which (given sufficient training data) estimate the relaxation timescales uncontaminated by discretizaiton error.  Note that there is no corresponding MSM for the OOM timescales.

\section*{Results}

\subsection*{A simple example of adaptive seeding estimation error}

To illustrate some of estimation errors that can arise with adaptive seeding, we first consider a simple model of 1D diffusion over a potential energy surface $V(x)$.  The continuous time evolution of a population distribution $p(x,t)$ is analytically described by the Fokker-Planck equation,
\begin{equation}
\frac{\partial}{\partial t}p(x,t) = D \frac{\partial^2}{\partial x^2}p(x,t) + \mu \frac{\partial}{\partial x}\big[ p(x,t) \nabla V(x) \big] 
\end{equation}
where $D$ is the diffusion constant and $\mu = D/k_BT$ is the mobility.  Here $x$ and $t$ are unitless values with $D$ set to unity and the energy function $u(x)$ in units of $k_BT$.

\begin{figure*}[ht!]
\includegraphics[width=0.5\paperwidth]{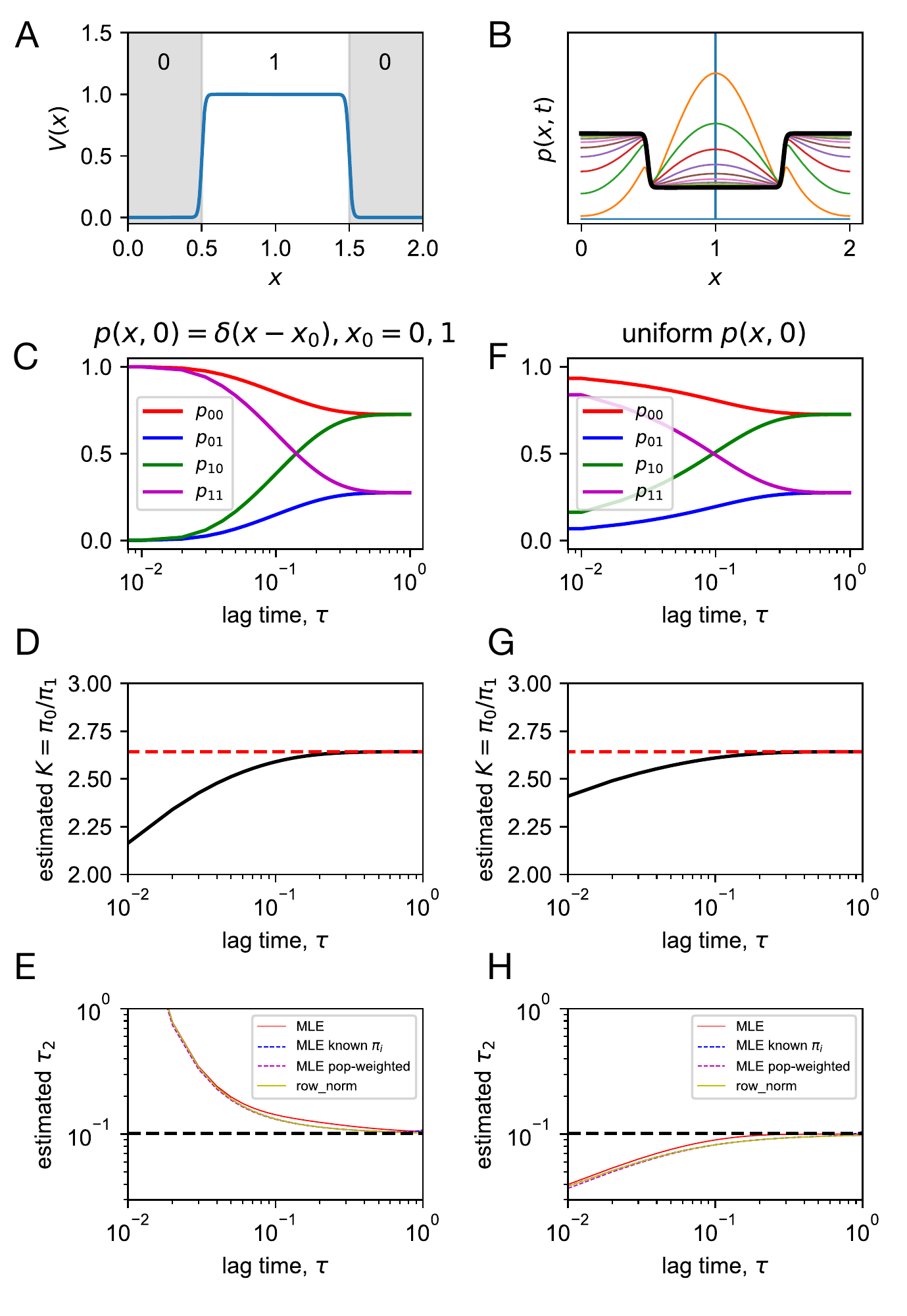}
\caption{Adaptive seeding in a simple two-state system. (A) The potential energy surface $V(x)$, annotated with state indices. (B) Propagation of an initial distribution $p(x,0) = \delta(x-1)$ (blue), plotted at increments $\Delta t$ = 0.1, up to $t=1$. Starting from an initial distribution $p(x,0) = \delta(x-x_0)$ using $x_0 = 0,1$, shown are estimated quantities as a function of lag time $\tau$: (C) transition probabilities $p_{ij}$, (D) equilibrium constant $K = \pi_0/\pi_1$, and (E) relaxation timescale $\tau_2$, as computed by various estimators. Starting from uniform distributions $p(x,0)$ over each state, shown are estimated quantities as a function of lag time $\tau$: (F) transition probabilities $p_{ij}$, (G) equilibrium constant $K = \pi_0/\pi_1$, and (H) relaxation timescale $\tau_2$, as computed by various estimators.}
\label{fig:fokker}
\end{figure*}

Here we consider a ``flat-bottom'' landscape, over a range $x \in (0,2)$ with periodic boundaries, described by
\begin{equation}
V(x) = \frac{e^{-b(|x-1|-0.5)}}{1 + e^{-b(|x-1|-0.5)}}.
\end{equation}
where $b = 100$.  The landscape is partitioned into two states: $x$ is assigned to state 0 if $x < 0.5$ or $x \geq 1.5$, and assigned to state 1 otherwise (Figure \ref{fig:fokker}A). In the limit of $b \rightarrow \infty$, the free energy difference between the states is exactly 1 ($k_BT$), with an equilibrium constant of $K = e \approx 2.718$.  When $b=100$, $K \approx 2.643$.  This value of $b$ was chosen to help the stability of numerical integration (performed at a resolution of $\Delta x = 2.5 \times 10^{-3}$ and $\Delta t = 5 \times 10^{-7}$). 

To emulate adaptive seeding, we propagate the continuous dynamics starting from an initial distribution $p(x,0)$ centered on state $i$, and compute the transition probabilities $p_{ij}$ from state $i$ to $j$ from the evolved density after some lag time $\tau$ (Figure \ref{fig:fokker}B). 

We first consider a seeding strategy where simulations are started from the center of each state, which we model by propagating density from an initial distributions $p(x,0) = \delta(x)$ to compute $p_{00}$ and $p_{01}$ as a function of $\tau$, and from $p(x,0) = \delta(x-1)$ to compute $p_{10}$ and $p_{11}$ (Figure \ref{fig:fokker}C).  By $t=1$, the population has reached equilibrium.  By detailed balance, the estimated equilibrium constant is $\hat{K} = p_{10}/p_{01}$, which systematically 
\textit{underestimates} the true value (Figure \ref{fig:fokker}D).  This is because the initial distribution is non-equilibrium, in the direction of being too uniform.  There are also errors in estimating the relaxation timescales.   The different estimators described above (MLE, MLE with known equilibrium populations $\pi_i$, population-weighted MLE, and row-normalized counts) were applied to estimate the implied timescale $\tau_2$ (Figure \ref{fig:fokker}E).  For this simple two-state system, all of the estimates are very similar to the analytical result $\tau_2 =  -\tau/\ln (1-p_{10}-p_{01})$ (with the exception of MLE, which yields $\tau_2$ estimates that are about 9\% larger when $t = 0.1$), which approaches the true value ($\tau_2 = 0.1013$) as the lag time increases.  Until the population density becomes equilibrated, however, the computed value of $\tau_2$ is \textit{overestimated}.  This is because at early times, outgoing population fluxes are underestimated.

We also consider a seeding strategy where simulations are started from randomly from inside each state, which we model by propagating density from initially uniform distributions across each state (Figure \ref{fig:fokker}F). As before, we find that the equilibrium constant is underestimated, again, because the initial seeding is out-of-equilibrium  (Figure \ref{fig:fokker}G).  The estimate is somewhat improved, however, by the ``pre-equilibration'' of each state.   In this case, the implied timescale $\tau_2$ is \textit{underestimated} at early times  (Figure \ref{fig:fokker}H).  This is can be understood as arising from the \textit{overestimation} of outgoing population fluxes, as not enough population has yet reached the ground state (state 0).

From these two scenarios, we can readily understand that the success of estimating both kinetics and thermodynamics from adaptive seeding greatly depends on how well the initial seeding distribution approximates the equilibrium distribution.  Moreover, we can see that timescale estimates can be either under-estimated or over-estimated, depending on the initial seeding distribution.
It is important to note that these errors arise primarily because of non-equilibrium sampling, rather than finite sampling error or discretization error.\cite{Prinz:2011id}

\subsection*{Seeding of a 1-D potential energy surface}

We next consider the following two-well potential energy surface, as used by Stelzl et al.\cite{Stelzl:2017gra}:  $U(x) = -\frac{2k_BT}{0.596} \ln [ e^{-2(x-2)^2-2} + e^{-2(x-5)^2} ] $
for $x \in [1.5,5.5]$, and $k_BT = 0.596$ kcal $\cdot$ mol$^{-1}$.  The state space is uniformly divided into 20 states of width 0.2 to calculate discrete-state quantities. Diffusion on the 1-D landscape is approximated by a Markov Chain Monte Carlo (MCMC) procedure in which new moves are translations randomly chosen from $\delta \in [-0.05, +0.05]$ and accepted with probability $\min(1, \exp( -\beta [U(x+\delta)-U(x)]))$, i.e. the Metropolis criterion.

\begin{figure}[ht!]
    \includegraphics[width=0.9\columnwidth]{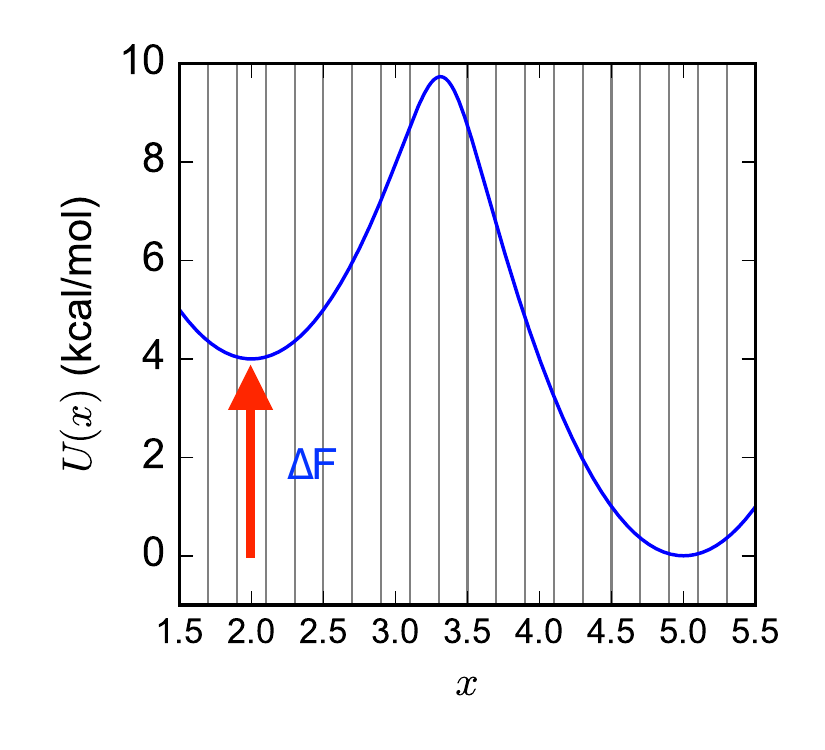}
    \caption{A 1-D two-state potential was used to test the performance of MSM estimators on seeding trajectory data.  Vertical lines denote the twenty uniformly-spaced discrete states}.
    \label{fig:1D-potential-v2}
\end{figure}

As a standard against which to compare estimates from seeding trajectories, we used a set of very long trajectories to construct an optimal MSM model using the above state definitions.   To estimate the  ``true'' relaxation timescales of the optimal model, we generated long MCMC trajectories of $10^9$ steps, sampling from a series of scaled potentials $U^{(\lambda)}(x) = \lambda U(x)$ for $\lambda \in [0.5, 0.6, 0.7, 0.8, 0.9, 1.0]$. For each $\lambda$ value, 20 trajectories were generated, with half of them starting from $x=2.0$ and the other half starting from $x=5.0$, resulting in a total of 120 trajectories.  The DTRAM estimator of Wu et al.\cite{Wu:2014jy} was used to estimate the slowest relaxation timescale as 9.66 ($\pm$ 1.37) $\times$ $10^6$ steps, using a lag time of 1000 steps.

We next generated adaptive seeding trajectory data for this system.  We limited the lag time to $\tau =$ 100 steps, and generated $s$ trajectories ($5\leq s \leq 1000$) of length $n\tau$ (for $n=10,100,1000$ by initiating MCMC dynamics from the center positions of all twenty bins.  The resulting data consisted of $20 \times s \times n$ transition counts between all states $i$ and $j$ in lag time $\tau$, stored in a 20 $\times$ 20 count matrix of entries $c_{ij}$.  From these counts, estimates of the transition probabilities, $p_{ij}$ were made. 

Estimates of the slowest MSM implied timescales were computed as $\tau_2 = -\tau/\ln \mu_2$, where $\mu_2$ is the largest non-stationary eigenvalue of the matrix of transition probabilities.  Estimates of the free energy difference $\Delta F$ between the two wells were computed as $-k_BT \ln (\pi_L/(1-\pi_L))$ where $\pi_L$ is the estimated total equilibrium population of MSM states with $x<3.3$. To estimate uncertainties in $\ln \tau_2$ and $\Delta F$, twenty bootstraps were constructed by sampling from the $s$ trajectories with replacement.

\begin{figure*}
\includegraphics[width=1.7\columnwidth]{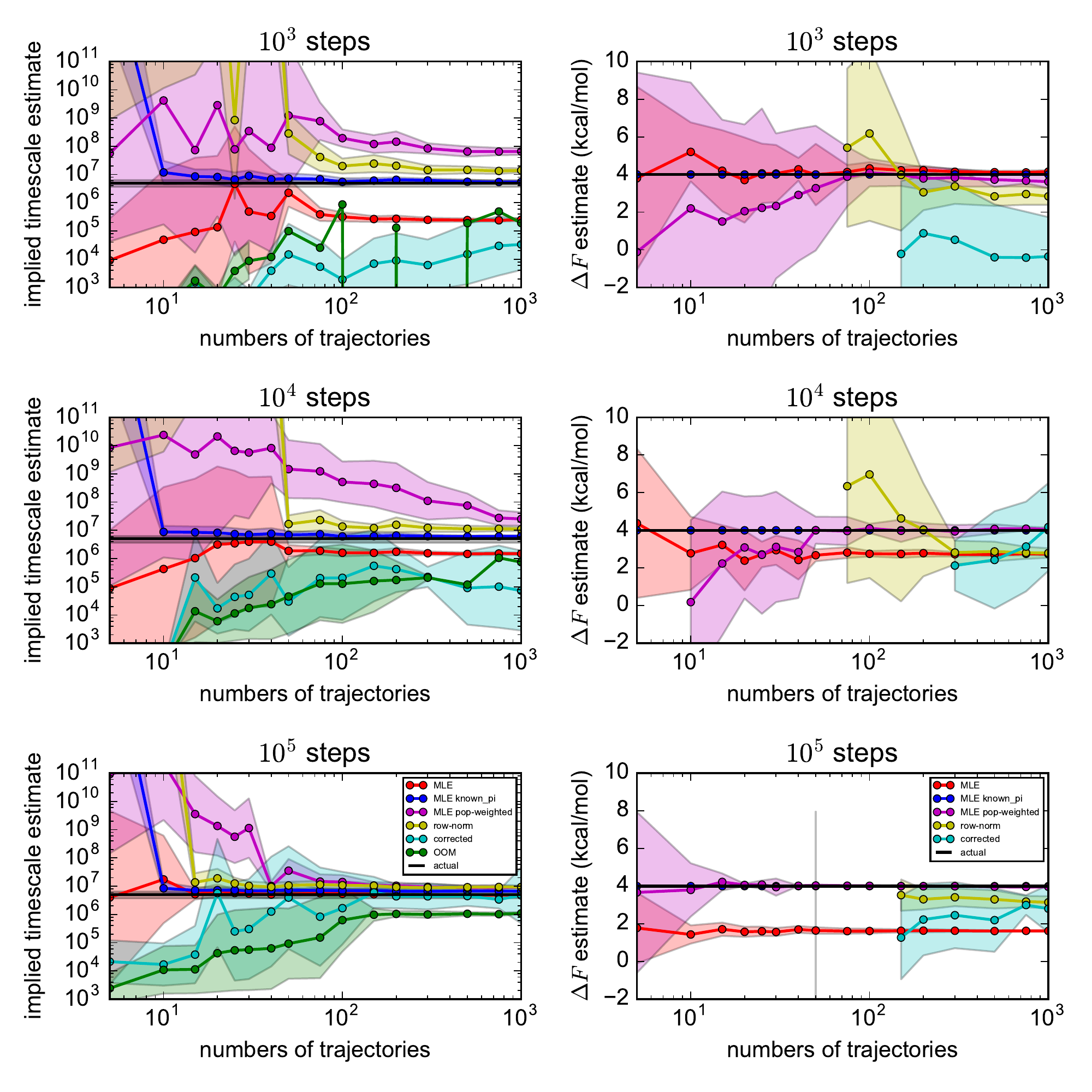}
\caption{\label{fig:toy-estimation}Estimates of the slowest implied timescales (left) and two-well free energy differences (right), shown as a function of number of seeding trajectories of various lengths. Shown are results for: (red) MLE, (blue) MLE with known equilibrium populations, (magenta) MLE with population-weighted trajectory counts, (yellow) row-normalized counts, (cyan) corrected MSMs, and (green) OOM timescales. Black horizontal lines denote the highly converged estimates from DTRAM.  Shaded bounds represent uncertainties estimated from a bootstrap procedure.}
\end{figure*}

Results for five different estimators are shown in Figure \ref{fig:toy-estimation}.  In all cases, timescale estimates improve with larger numbers of seeding trajectories, and greater numbers of steps in each trajectory.  From these results, one can clearly see that MLE (red lines in Figure \ref{fig:toy-estimation}), the default estimator in software packages like PyEMMA\cite{Scherer:2015jb} and MSMBuilder\cite{Harrigan:2017gw}, consistently underestimates the implied timescale, although this artifact becomes less severe as more trajectory data is incorporated.  At the same time, MLE estimates of equilibrium populations become systematically \textit{worse} with more trajectory data.  This is precisely because uniform seeding produces a biased, \textit{non-equilibrium} distribution of sampled trajectories, but the MLE assumes that trajectory data is drawn from \textit{equilibrium}. In contrast, the MLE with known equilibrium populations (blue lines in Figure \ref{fig:toy-estimation}) accurately predicts implied timescales with much less trajectory data, underscoring the powerful constraint that detailed balance provides in determining rates.  Of all the estimators, this method is the most accurate, but relies on having very good estimates of equilibrium state populations \textit{a priori}.

The row-normalized counts estimator (yellow lines in Figure \ref{fig:toy-estimation}) estimates timescales well, provided that a sufficient number of observed transitions must be sampled to obtain proper estimates.  This is necessary to achieve connectivity and to ensure that the transition matrix is at least approximately reversible. However, the free energy differences estimated by row-normalized counts are highly uncertain, and in the limit of large numbers of sufficiently long trajectories ($10^5$ steps), estimates of free energies become systematically incorrect due the the statistical bias from seeding. In comparison, the MLE with population-weighted trajectory counts (magenta lines in Figure \ref{fig:toy-estimation}) is able to make much more accurate estimates of free energies with less trajectory data.  We note that while here we are reweighting the transition counts with the known populations, the results are similar if reasonable approximations of the state populations are used (data not shown). Compared to the row-normalized counts estimator, implied timescales are slower to converge with more trajectory data, but are comparable when the seeding trajectories are sufficiently long ($10^5$ steps).  With sufficiently long trajectories ($10^5$ steps), the MLE  timescales are better estimated than either the row-normalized counts estimator or population-weighted trajectory counts, but MLE is flawed because the free energies are so biased from the seeding.  The population-weighting trajectory counts estimator avoids this artifact to give good estimates of free energies.

\begin{figure}[ht!]
    \includegraphics[width=0.85\columnwidth]{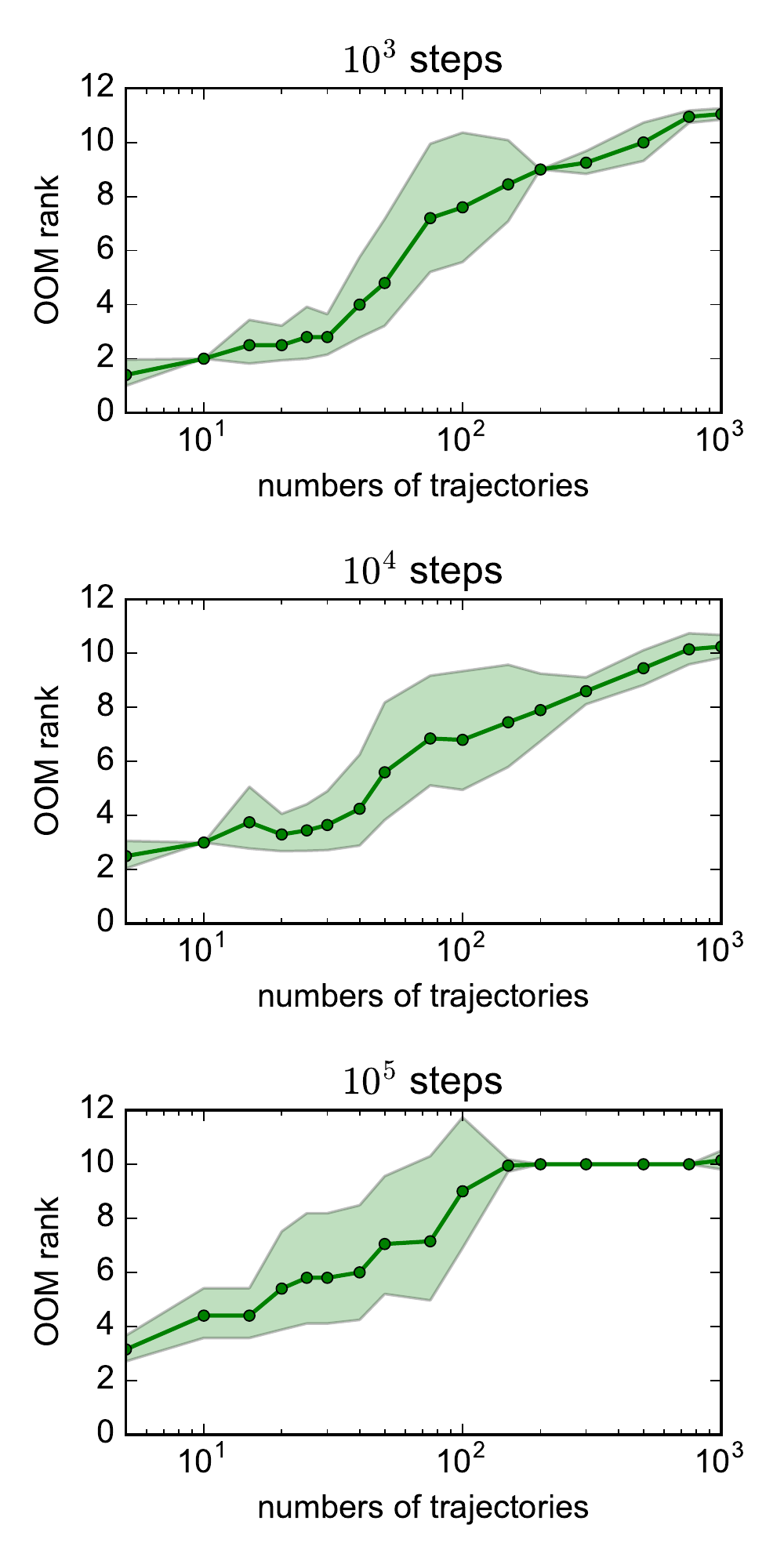}
    \caption{The average OOM rank increases with the number of trajectories.  Shaded bounds represent uncertainties estimated from a bootstrap procedure.}.
    \label{fig:rankM}
\end{figure}

The ``corrected'' and OOM timescales from the OOM estimator consistently underestimate the slowest implied timescale for this test system, although these estimates improve with greater numbers and lengths of trajectories (cyan and green lines in Figure \ref{fig:toy-estimation}).  The OOM estimator is susceptible statistical noise, as evidenced by the average OOM model rank selected by the PyEMMA implementation, which increases from 2 to 10 as the number of trajectories are increased (Figure \ref{fig:rankM}).  Like the row-normalized counts estimator, the OOM estimator gives poorly converged estimates of two-well free energy differences.

\subsection*{Adaptive seeding of folding landscapes for WW Domain and NTL9(1-39)}

For studying protein folding, an adaptive seeding strategy may be useful for several reasons. For one, generating ensembles of short trajectories may simply be a practical necessity, due to the limited availability of special-purpose hardware to generate ultra-long trajectories.  Adaptive seeding may also be able to efficiently leverage an existing MSM (perhaps built from ultra-long trajectory data) to predict perturbations to folding by mutations, or to examine the effect of different simulation parameters like force field or temperature. 

How can we best utilize adaptive seeding of MSMs built from ultra-long trajectory data to make good estimates of both kinetics and thermodynamics?  To address this question, here we test the performance of various MSM estimators on the challenging task of estimating MSMs from adaptive seeding of protein folding landscapes.  The folding landscapes come from MSMs built from ultra-long reversible folding trajectories of fast-folding mini-proteins WW domain and NTL9(1-39).

WW domain is a 35-residue protein signaling domain with a three-stranded $\beta$-sheet structure that binds to proline-rich peptides.  Its folding kinetics and thermodynamics have been studied extensively by time-resolved spectroscopy,\cite{Nguyen:2003gw,Jager:2006ga,Liu:2008fs,Dave:2016joa,Liu:2008kj} and its folding mechanism has been probed extensively using molecular simulation studies.\cite{Gao:2017kna,Best:2013hua,Piana:2011gma} Moreover, an impressively large number of site-directed mutants of WW domain have been constructed to investigate folding mechanism and discover fast-folding variants, including the Fip mutant (Fip35) of human pin1 WW domain, for which mutations in first hairpin loop region resulting in a fast folding time of $\sim$13.3 $\mu$s at 77.5 $^{\circ}$C.\cite{Jager:2006ga,Nguyen:2005ec}  Further mutation of the the second hairpin loop of Fip35 produced an even faster-folding variant, called GTT, in which the native sequence Asn-Ala-Ser (NAS) was replaced with Gly-Thr-Thr (GTT), resulting in a a fast relaxation time $\sim$4.3 $\mu$s at 80 $^{\circ}$C.\cite{Piana:2011gma}   

NTL9(1-39) domain is a 39-residue truncation variant of the N-terminal domain of ribosomal protein L9, whose folded state consists of an $\alpha$-helix and three-strand $\beta$-sheet. NTL9(1-39), which has a folding relaxation time of $\sim$1.5 ms at 25 $^{\circ}$C,\cite{Horng:2003gf} has been extensively probed by both experimental and computational studies.\cite{Cho:2005is,Voelz:2010hsa,Schwantes:2013bpa,Cho:2014fla, Baiz:2014gm} The K12M mutant of NTL9(1-39) is a fast-folding variant with a folding timescale of $\sim$700 $\mu$s at 25 $^{\circ}$C.\cite{Horng:2003gf,Cho:2004dp}

\begin{figure*}[ht!]
\includegraphics[width=0.8\paperwidth]{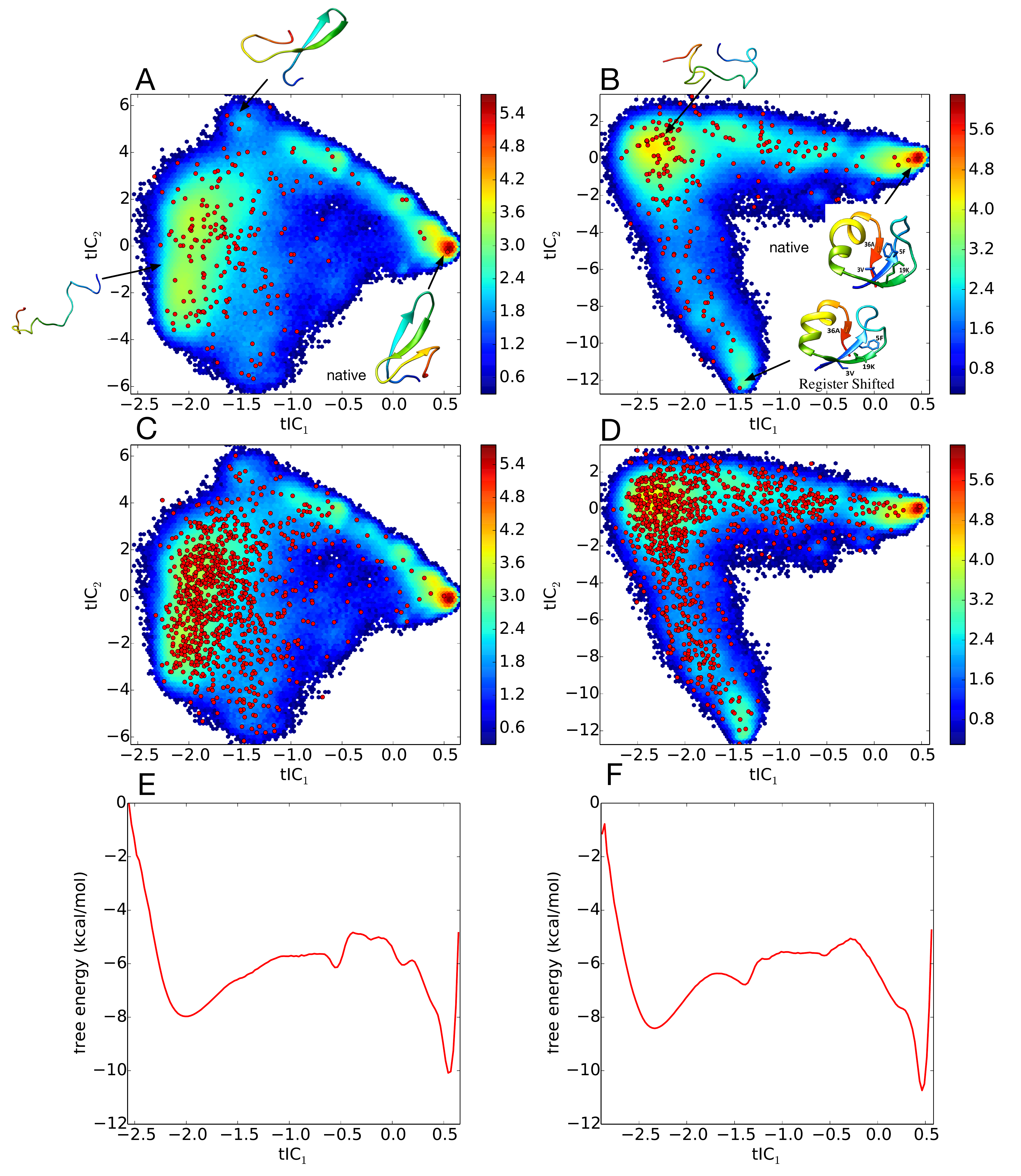}
\caption{200-state MSMs of (A) GTT WW domain and (B) K12M NTL9(1-39), built from ultra-long folding trajectory data, are shown projected onto the two largest tICA components. Red circles show the locations of the MSM microstates.  The heat map shows a density plot of the raw trajectory data, with color bar values denoting the natural logarithm of histogram counts. 1000-state MSMs are shown for (C) GTT WW domain, and (D) K12M NTL9(1-39). (E) and (F) show corresponding 1-D free energy profiles along the first tICA component, calculated from histogramming the trajectory data in bins of width 0.025.}
\end{figure*}

\begin{figure*}[ht!]
\includegraphics[width=0.6\paperwidth]{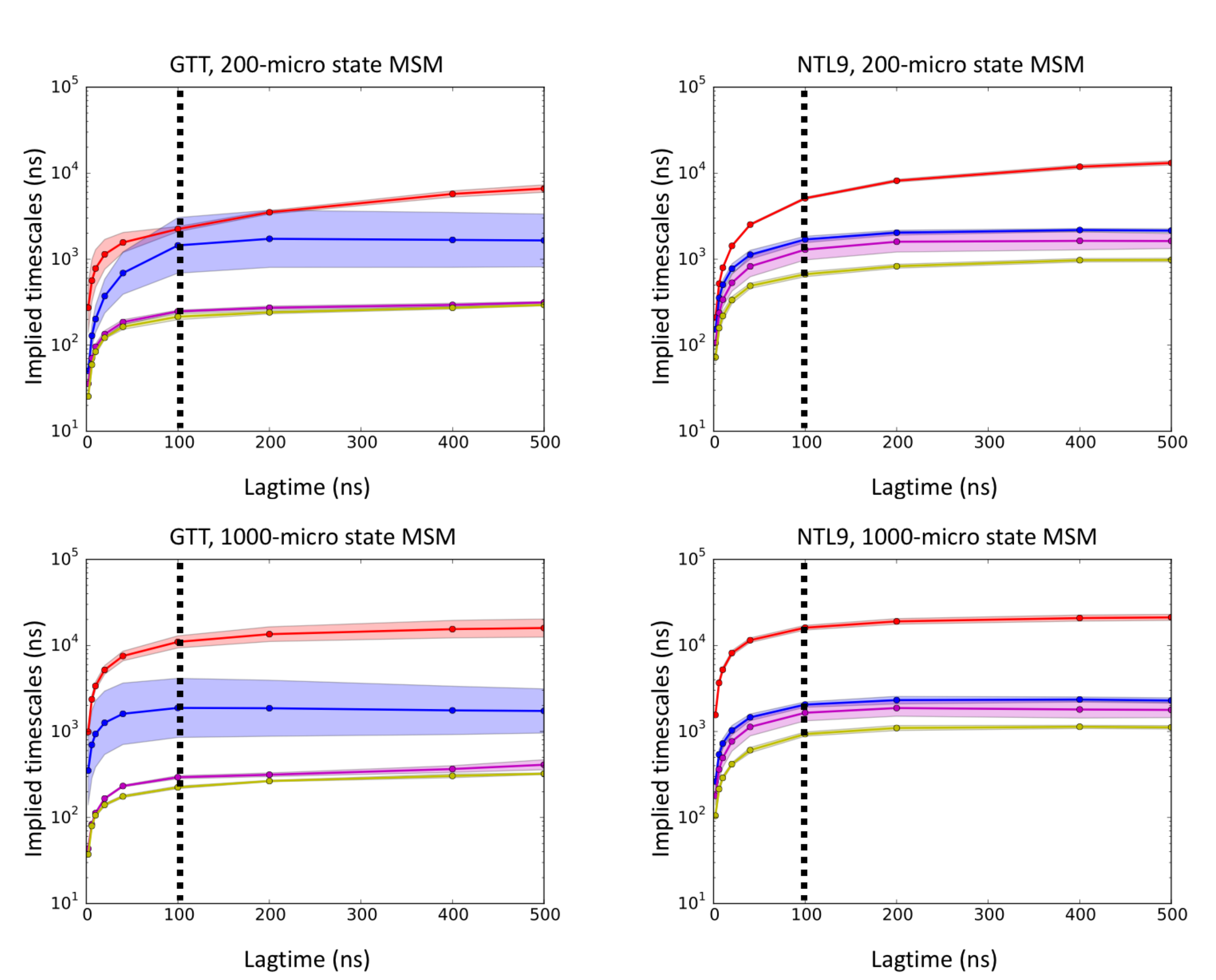}
\caption{Implied timescale plots for 200-state MSMs of (A) GTT WW domain and (B) K12M NTL9(1-39), and 1000-state MSMs of (C) GTT WW domain and (D) K12M NTL9(1-39).  Uncertainties (shaded areas) for GTT and NTL9 were estimated using 5-fold and 8-fold leave-one-out bootstraps, respectively. The dotted vertical line marks the lag time of 100 ns chosen to constuct the MSMs.}
\end{figure*}

\subsubsection*{Markov State Model construction}

Molecular simulation trajectory data for GTT WW domain and K12M NTL9(1-39) were provided by Shaw et al.,\cite{LindorffLarsen:2011gla}  Our first objective was to construct high-quality reference MSMs from the available trajectory data, to be used as benchmarks against which we could compare adaptive seeding results.

Two independent ultra-long trajectories of 651 and 486 $\mu$s, performed at 360K using the CHARMM22* force field, were used to construct MSMs of GTT WW domain, as previously described by Wan and Voelz.\cite{Wan:2016fz}   The MSM was constructed by first projecting the trajectory data to all pairwise distances between C$_{\alpha}$ and C$_{\beta}$ atoms, performing dimensionality reduction via  tICA (time-lagged independent component analysis),\cite{Schwantes:2013bp, PerezHernandez:2013tt} and then clustering in this lower-dimensional space using the $k$-centers algorithm to define a set of metastable states.  A 1000-state MSM for GTT WW domain was constructed using a lag time of 100 ns.  The generalized matrix Raleigh quotient (GMRQ) method\cite{McGibbon:2015gm} was used to choose the optimal number of states (1000) and number of independent tICA components (eight).   The MSM gives an estimated folding relaxation timescale of 10.2 $\mu$s, which is comparable to the folding time of $21 \pm 6$ $\mu$s estimated from analysis of the trajectory data by Lindorff-Larsen et al.\cite{LindorffLarsen:2011gla}  It is also comparable to the 8 $\mu$s timescale estimate from a three-state MSM model built using sliding constraint rate estimation by Beauchamp et al.\cite{Beauchamp:2012kp}  As described below, since seeding trajectory data for a 1000-state MSM was too expensive to analyze using the OOM estimator, we additionally built a 200-state model, using the same methods, which gives an estimated folding timescale of 2.3 $\mu$s.

Four trajectories of the K12M mutant of NTL9(1-39), of lengths 1052 $\mu$s, 990 $\mu$s, 389 $\mu$s, and 377 $\mu$s, simulated at 355 K using the CHARMM22* force field, were provided by Shaw et al.\cite{LindorffLarsen:2011gla}.  Choosing a lagtime of 200 ns, we constructed an MSM using the same methodology as GTT WW domain described above, resulting in a 1000-state model utilizing 6 tICA components which gives a folding timescale estimate of $\sim$18 $\mu$s.  Although this is a faster timescale than the experimentally measured value at 25 $^{\circ}$C, it accurately reflects the timescale of folding events observed in the trajectory data (at 355 K), and is similar to folding timescales measured by T-jump 2-D IR spectroscopy at 80 $^{\circ}$C ($\sim$26.4 $\mu$s).\cite{Baiz:2014gm} We also note that previous analysis of the K12M NTL9(1-39) trajectory data has given similar timescale estimates. Analysis of the trajectory data by Lindorff-Larsen et al. yielded an estimate of $\sim$29 $\mu$s,\cite{LindorffLarsen:2011gla} a three-state MSM model built using sliding constraint rate estimation by Beauchamp et al. gives an estimate of $\sim$16 $\mu$s,\cite{Beauchamp:2012kp} and an MSM by Baiz et al. gives a timescale estimate of 18 $\mu$s.\cite{Baiz:2014gm}  As we did for GTT WW domain, and using the same methods, we additionally built a 200-state MSM, which gives an estimated folding timescale of 5.2$\mu$s.

\subsubsection*{Estimated folding rates and equilibrium populations from reseeding simulations}

To emulate data sets that would be obtained by adaptive reseeding, we randomly sampled segments of the reference trajectory data that start from each metastable state.  An advantage of emulating reseeding trajectory data in this way is the ability to recapitulate the original model in the limit of long trajectories. The full set of reference trajectory data ($\sim$1.1 ms) for GTT WW domain comprised 12 reversible folding events, while the full set for K12M NTL9(1-39) ($\sim$ 2.8 ms) comprised 14 reversible folding events. 

To test the performance of different estimators on reseeding trajectory data, we sampled either 5 or 10 seeding trajectories of various lengths from each MSM state.  By design, the length of the reseeding trajectories were limited to 500 ns for the 200-state MSMs, and 250 ns for the 1000-state MSMs, because trajectories longer than this length result in datasets that exceed the total amount of reference trajectory data.  For reseeding 1000-state MSMs with 10 trajectories, the trajectory lengths were limited to 100 ns for GTT, and 200 ns for NTL9, so as not to exceed the total amount of original trajectory data.

For each set of emulated reseeding trajectory data, MSMs were constructed using various estimators and compared to the reference MSM. Implied timescale estimates were computed using a lag time of 40 ns, to facilitate the analysis of seeding trajectories as short as 50 ns. For the MLE estimator with population reweighting, and the MLE estimator with known populations, we used the equilibrium populations estimated from the reference MSMs.  In cases where the row-normalized counts estimator fails at large lag times due to the disconnection of states, a single pseudo-count was added to elements of the transition count matrix that are nonzero in the reference MSM.


\begin{figure*}[ht!]
\includegraphics[width=0.55\paperwidth]{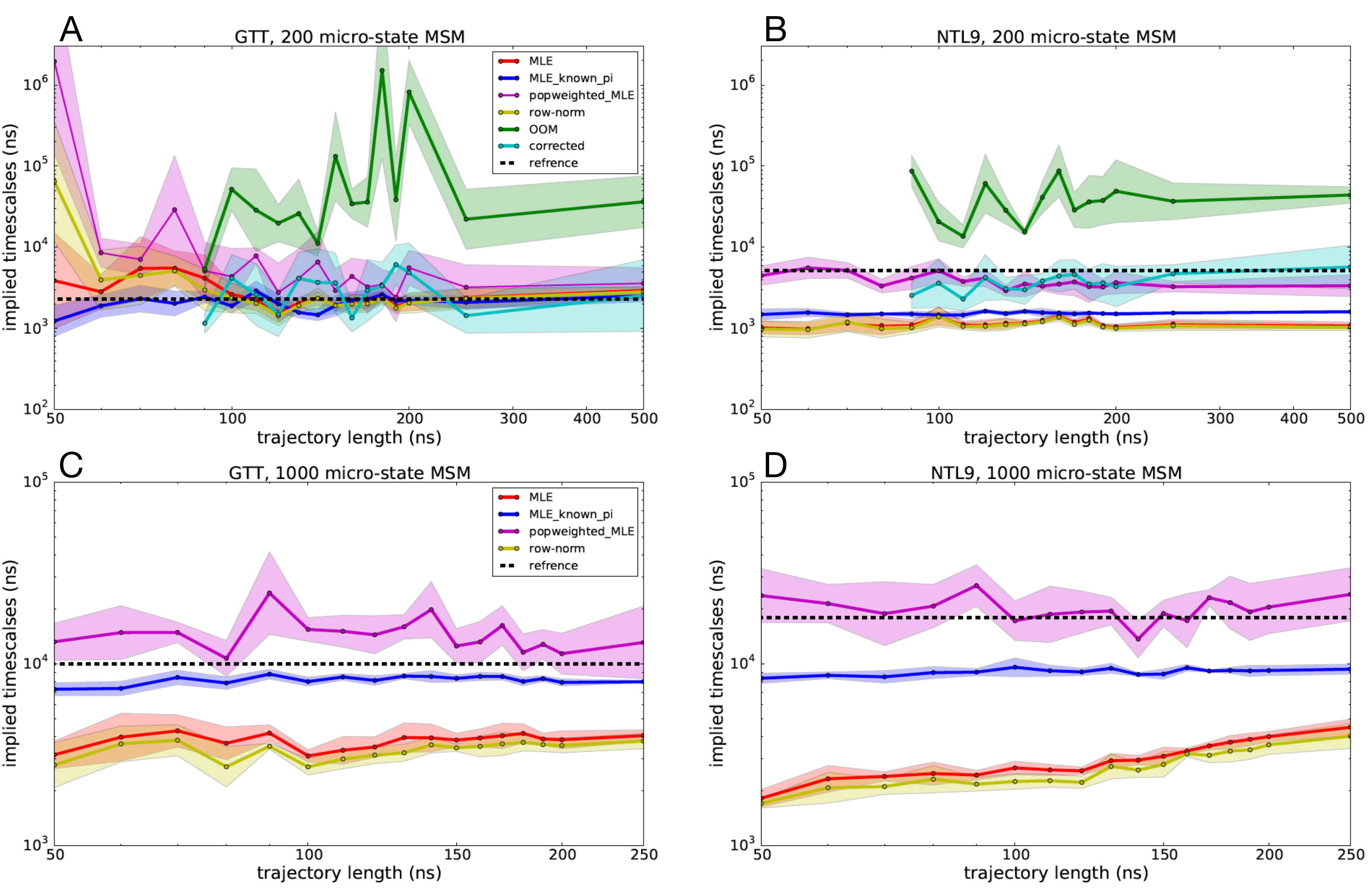}
\caption{Slowest implied timescales predicted by various MSM estimators as a function of seeding trajectory length. Estimated slowest timescales are shown for 200-microstate MSMs of (A) GTT WW domain and (B) K12M NTL9(1-39), and 1000-microstate MSMs of (C) GTT WW domain and (D) K12M NTL9(1-39). Adaptive seeding data was generated using five independent trajectories from each microstate. Uncertainties were computed using a bootstrap procedure.}
\label{fig:all-atom-timescales-5seeds}
\end{figure*}

\begin{figure*}[ht!]
\includegraphics[width=0.55\paperwidth]{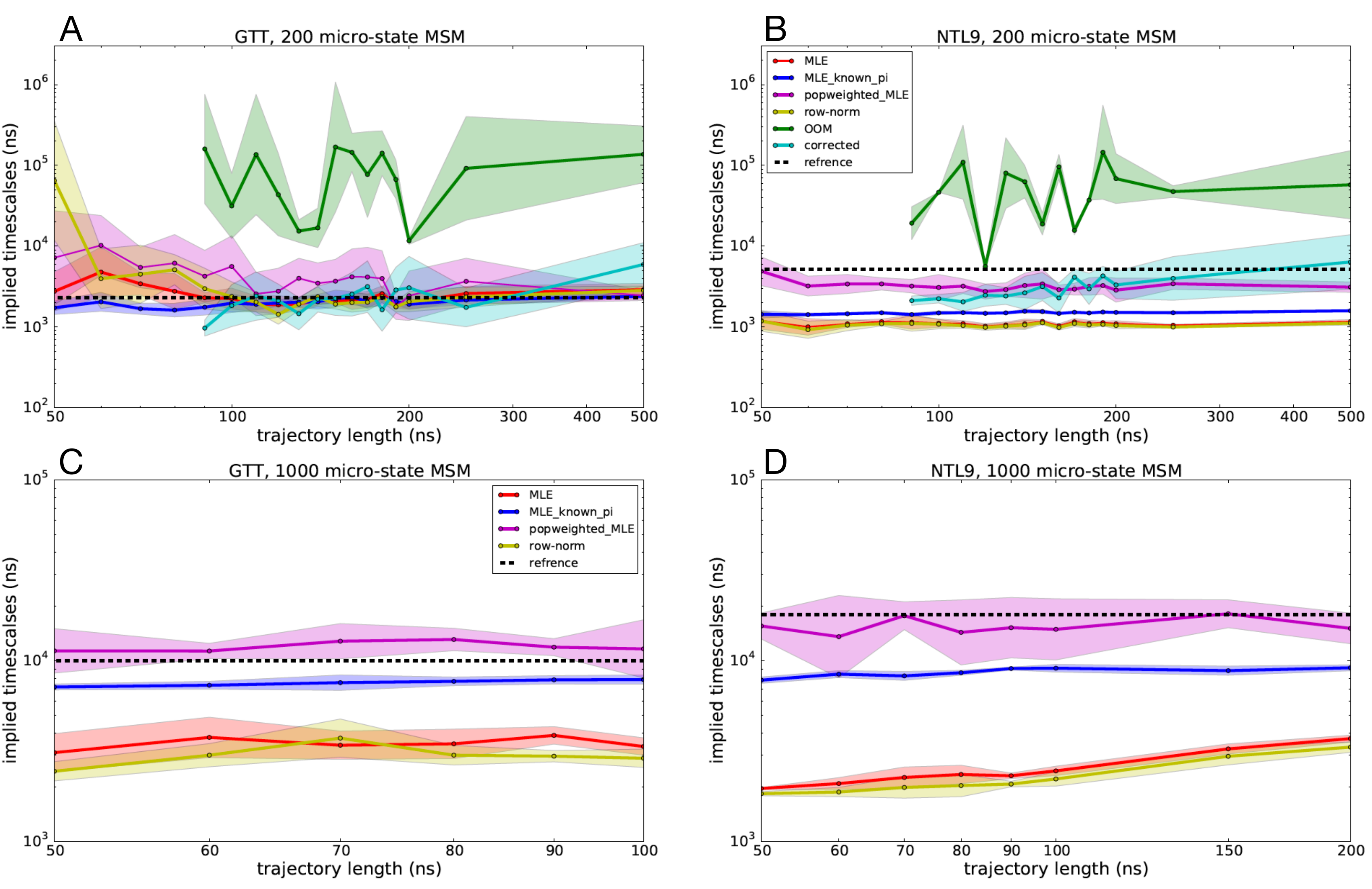}
\caption{Slowest implied timescales predicted by various MSM estimators as a function of seeding trajectory length. Estimated slowest timescales are shown for 200-microstate MSMs of (A) GTT WW domain and (B) K12M NTL9(1-39), and 1000-microstate MSMs of (C) GTT WW domain and (D) K12M NTL9(1-39). Adaptive seeding data was generated using ten independent trajectories from each microstate. Trajectory lengths longer than 100 ns for GTT WW domain and 200 ns for K12M NTL9(1-39) result in trajectory datasets larger than the reference model, and are not shown. Uncertainties were computed using a bootstrap procedure. }
\label{fig:all-atom-timescales-10seeds}
\end{figure*}

\paragraph{Estimates of slowest implied timescales.}

We applied different estimators to the reseeding trajectory data generated for 200-state and 1000-state MSMs of GTT WW domain and K12M NTL9(1-39), and compared the predicted slowest implied timescales to the reference MSMs. Similar implied timescale estimates were obtained using five independent reseeding trajectories initiated from each state (Figure \ref{fig:all-atom-timescales-5seeds}) and ten independent reseeding trajectories (Figure \ref{fig:all-atom-timescales-10seeds}).

Across most of the estimators tested, implied timescale estimates are statistically well-converged using trajectories of 100 ns or more.  For the MLE-based estimators in particular (MLE, MLE with known populations, MLE with population reweighting), the estimates have small uncertainties and are not strongly sensitive to trajectory lengths tested. This suggests that while there is enough seeding data to combat statistical sampling error, there remain systematic errors in estimated timescales, mainly due to the estimation method used.

The MLE performs well for the 200-state MSMs of GTT, but consistently underestimates the slowest implied timescale in all other MSMs.  As seen with our tests above with the 1-D two-state potential, this error is likely due to using a detailed balance constraint with statistically biased seeding trajectory data. The MLE with known populations is much more accurate at estimating the slowest implied timescale, and with less uncertainty. For the NLT9 MSMs, however, MLE with known populations systematically underestimates the slowest implied timescale, again suggesting the biased seeding trajectory data has an influence despite the strong constraints enforced by the estimator. MLE with population reweighting has more uncertainty than the other MLE-based estimators, but overall is arguably the most accurate estimator across all the MSM reseeding tests.

The OOM-based estimates of the slowest implied timescale (which could only be performed for the computationally tractable 200-state MSMs) required seeding trajectory data of least 90 ns, in order to have sufficient trajectory data to perform the bootstrapped signal-to-noise estimation, from which the OOM rank is selected.  While the predicted OOM timescales overestimate the relaxation timescales of the 200-state MSMs of both GTT and NTL9 systems, the timescale estimates from the corrected MSMs more closely recapitulate the reference MSMs timescales, with accuracy comparable to the MLE with population reweighting, especially for seeding trajectories over 200 ns.

The row-normalized counts estimator yields slowest implied timescales estimates accelerated compared to the reference MSM, similar to the MLE results.  This acceleration is greater than might be expected given our tests with the 1-D two-state potential, and may in part be due to the necessity of adding pseudo-counts to the transition count matrix to ensure connectivity.

\paragraph{Estimates of native-state populations.}

\begin{figure*}[ht!]
\includegraphics[width=0.55\paperwidth]{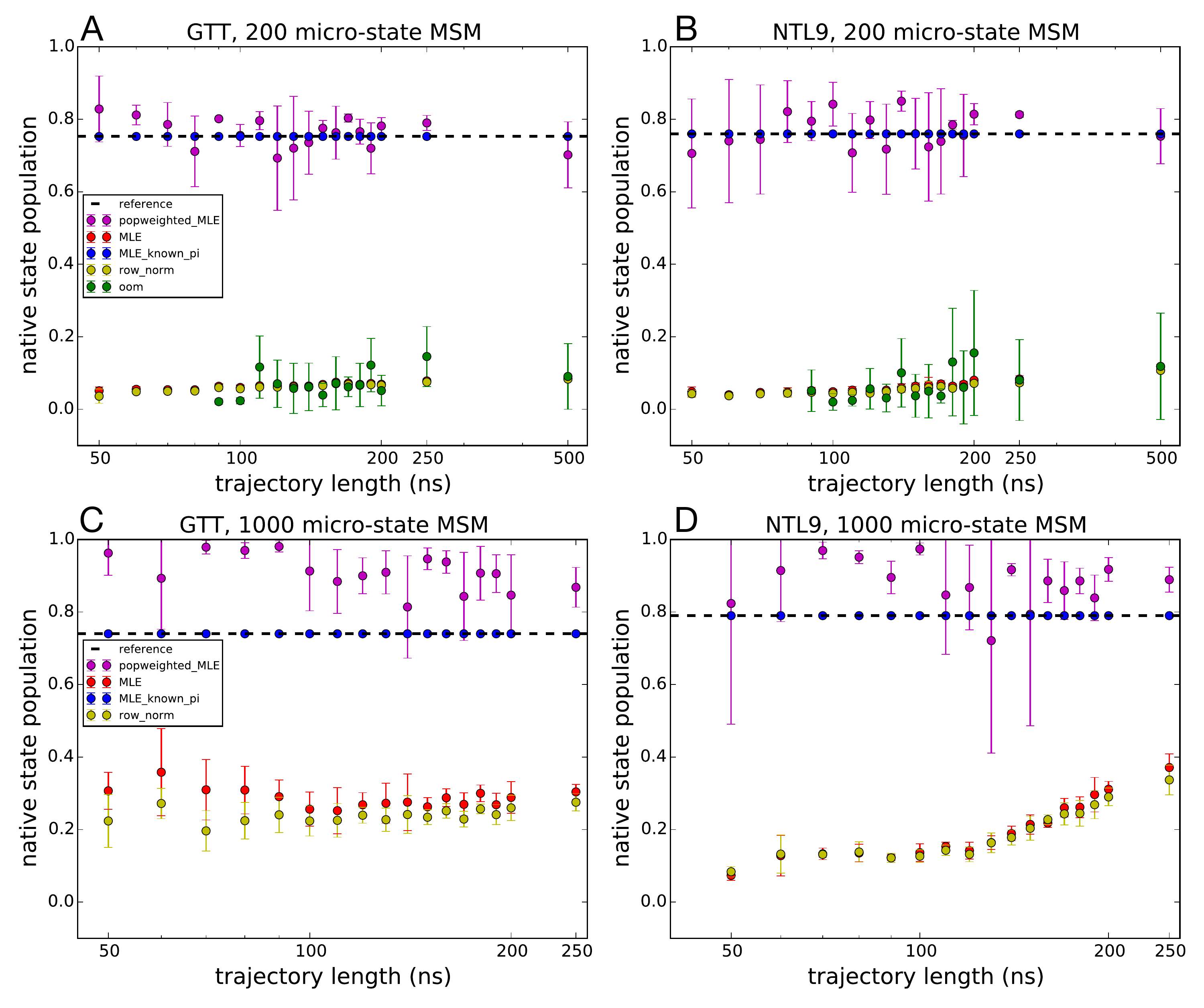}
\caption{Native state populations predicted by various MSM estimators as a function of seeding trajectory length. Estimated populations are shown for 200-microstate MSMs of (A) GTT WW domain and (B) K12M NTL9(1-39), and 1000-microstate MSMs of (C) GTT WW domain and (D) K12M NTL9(1-39). Adaptive seeding data was generated using five independent trajectories from each microstate. Uncertainties were computed using a bootstrap procedure.}
\label{fig:all-atom-nativepops-5seeds}
\end{figure*}

\begin{figure*}[ht!]
\includegraphics[width=0.55\paperwidth]{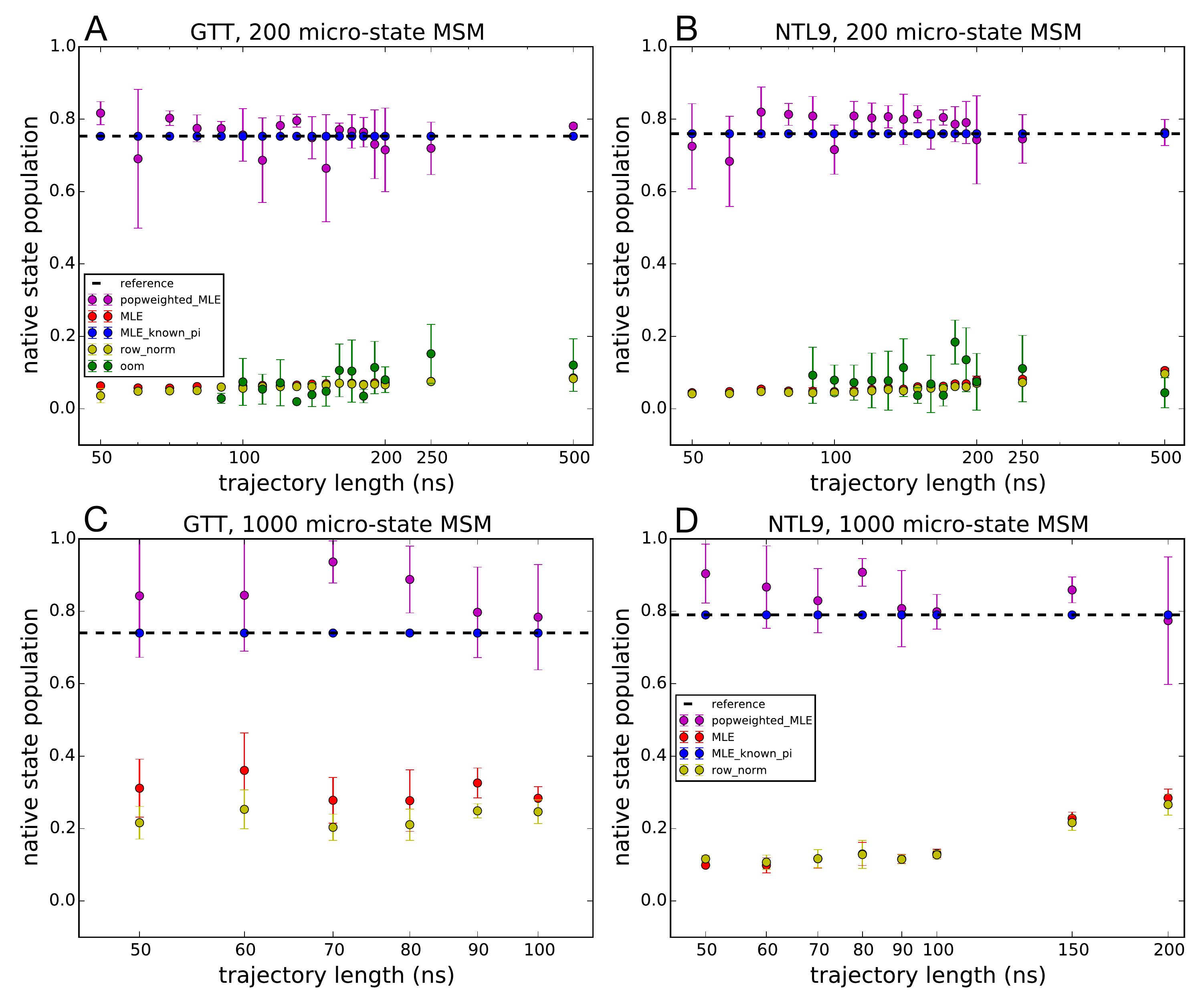}
\caption{Native state populations predicted by various MSM estimators as a function of seeding trajectory length. Estimated populations are shown for 200-microstate MSMs of (A) GTT WW domain and (B) K12M NTL9(1-39), and 1000-microstate MSMs of (C) GTT WW domain and (D) K12M NTL9(1-39). Adaptive seeding data was generated using ten independent trajectories from each microstate. Trajectory lengths longer than 100 ns for GTT WW domain and 200 ns for K12M NTL9(1-39) result in trajectory datasets larger than the reference model, and are not shown. Uncertainties were computed using a bootstrap procedure.}
\label{fig:all-atom-nativepops-10seeds}
\end{figure*}

We applied different estimators to the reseeding trajectory data generated for 200-state and 1000-state MSMs of GTT WW domain and K12M NTL9(1-39), and compared the predicted native-state populations to the reference MSMs. The native-state population was calculated as the equilibrium population of the MSM state corresponding to the native conformation.  Similar native-state populations were estimated using five independent reseeding trajectories initiated from each state (Figure \ref{fig:all-atom-nativepops-5seeds}) and ten independent reseeding trajectories (Figure \ref{fig:all-atom-nativepops-10seeds}). These estimates are compared to the native-state populations estimated from the reference MSMs: 75.3\% and 74.8\% for the 200-state and 1000-state MSMs of GTT WW domain, respectively, and 75.9\% and 79.4\% for the 200-state and 1000-state MSMs of K12M NTL9(1-39), respectively.

As we saw in our tests using a 1-D two-state potential, obtaining accurate estimates of the equilibrium populations from very short seeding trajectories is a challenging task.  In those tests, we found that the most accurate estimates of native-state populations come from estimators that utilize some \textit{a priori} knowledge of native-state populations. Here, the MLE with known populations recapitulates the native population exactly (by definition), while the MLE with population reweighting successfully captures the the native-state population in a 200 micro-state model of both GTT and NTL9, and slightly overestimates the native state population for the 1000-state MSMs. In contrast, the MLE, row-normalized counts, and OOM estimators all underestimate the folded populations.  With the exception of some slight improvement that begins to occur as the seeding trajectory length increases, these underestimates are independent of the trajectory length, indicating that that non-equilibrium seeding is to blame.  The native state populations are underestimated in these cases because there are many more MSM microstates belonging the unfolded-state basin of the folding landscape.

\section*{Discussion}

Markov State Model approaches have enjoyed great success in the last decade due to their ability to integrate ensembles of short trajectories.  In light of this success, the allure of adaptive sampling methods, which promise to leverage the focused sampling of short trajectories, is understandable.  Recent studies have shown that a judiciously-chosen adaptive sampling strategy can accelerate barrier-crossing, state discovery and pathway sampling.\cite{Hruska:2018ec,Zimmerman:2018jn}

Our work in this manuscript underscores that a key problem with the practical use of adaptive sampling algorithms is the statistical bias introduced by focused sampling, which makes difficult the unbiased estimation of both kinetics and thermodynamics. We have illustrated these problems in simple 1D diffusion models, and in large-scale all-atom simulations of protein folding, by testing the performance of various MSM estimators on adaptive seeding data generated these scenarios. In all cases, we find that estimators that incorporate some form of \textit{a priori} knowledge about equilibrium populations are able to correct for sampling bias to some extent, resulting in accurate estimates of both slowest implied timescales and equilibrium free energies.  Moreover, we have shown that an MLE estimator with population-weighted transition counts is a very simple way to counteract this bias, and achieve reasonably accurate results.

As pointed out by others previously,\cite{trendelkamp2016efficient} the success of such estimators is further demonstration of the importance of incorporating thermodynamic information into MSM estimates, given the powerful constraint that detailed balance provides. With this in mind, it is no surprise that new multi-ensemble MSM estimators like TRAM\cite{Wu:2016fk} and DHAMed\cite{Stelzl:2017gra} have been able to achieve more accurate results than previous estimators. Thus, while many existing adaptive sampling strategies focus sampling to refine estimated transition rates between states, we expect that improved adaptive sampling estimators may come from similar ``on-the-fly'' focusing that seeks to refine thermodynamic estimates as well.  

As adaptive sampling strategies and estimators continue to improve, we expect the more efficient use of ultra-long simulation trajectories combined with ensembles of short trajectories for adaptive sampling, especially to probe the effects of protein mutations, or different binding partners for molecular recognition.

\begin{acknowledgments}
The authors thank the participants of Folding@home, without whom this work would not be possible.  We thank D. E. Shaw Research for providing access to folding trajectory data. This research was supported in part by the National Science Foundation through major research instrumentation grant number CNS-09-58854 and National Institutes of Health grants 1R01GM123296 and NIH Research Resource Computer Cluster Grant S10-OD020095.
\end{acknowledgments}

%

\bibliography{seeding}

\newpage






\end{document}